# IR Reflection mechanism inside the annulus between two concentric cylindrical tubes

Khaled Mohamad[a], P. Ferrer[a,b]

[a]*School of Physics, Materials for Energy Research Group, University of the Witwatersrand, Johannesburg, 2001, South Africa.*
[b]*School of Physics, Mandelstam institute for Theoretical Physics, University of the Witwatersrand, Johannesburg, 2001, South Africa.*

A mathematical model was derived to calculate the IR reflection inside the annulus between two concentric cylindrical tubes, where the inner side of the outer cylinder is assumed to be coated with an IR reflected mirror. The mathematical model is implemented in a simulation code and experimentally validated. The experimental results of a system of two concentric cylindrical tubes operating with an IR reflected mirror on the inside of the outer cylinder are presented, and the results between the model and the simulation are compared. It is seen that the correspondence is encouragingly close (Chi-squared test p-values between 0.995 and 0.80), where the simulation underestimates the experimental performance.

## 1. Introduction

The applications of two concentric cylindrical tubes system are mostly used for heat exchange applications, especially for parabolic trough receiver unit. The Parabolic Trough Collector (PTC) technology are amongst the most mature commercially developed solar power technologies [1]. They operate by focusing solar radiation along a line onto a receiver unit which exchanges heat with a circulating heat transfer fluid (HTF). The heat gained by the HTF can be used either directly in thermal applications or for electricity production [2]. The receiver traditionally consists of an inner pipe coated with a selective coating and covered by a glass envelope. The air between the inner pipe and the glass envelope is evacuated to reduce convective and conductive heat transfer, making radiation the dominant heat loss mechanism.

Numerous works have investigated the thermal properties and heat retention of the receiver unit from various perspectives. Some concentrate on improving the uniformity of the thermal distribution of the metal inner pipe to reduce thermal stress and deformation [3][4].

The IR reflection mechanism inside the annulus of the PTC receiver unit is investigated by coating the inside glass cover around the inner pipe with a dielectric material that is transparent to the visible region of the solar spectrum and reflects well in the IR region. A coating of this type is referred to as a "hot mirror", and was first implemented for energy-efficient windows in automobiles and buildings [5] and for applications related to concentrating photovoltaics and thermophotovoltaics [6][7]. There are two general types of hot mirror films: a semiconducting oxide with a high doping level and a very thin metal film sandwiched between two dielectric layers (see [8][9][5] for more details). The thin metal film coating shows some unavoidable losses. Besides, the highly doped semiconducting oxide shows more advantages, i.e., Indium-Tin-Oxide (ITO). The hot mirror coating for a solar collector must meet some performance specifications. It needs to be highly transparent in the visible region and have high reflectivity in the IR region of the solar spectrum. Granqvist et al. [5] and Lampert et al. [8] focused on improving the transparency in the visible region and the reflectivity in the IR [8]. The advantage of applying a hot mirror coating to the glass cover as opposed to a selective coating onto the inner pipe lies in the fact that the glass cover is typically hundreds of degrees cooler than the inner pipe.

The effects of the hot mirror film have been modeled and studied previously. Grena [9] simulated the system including heat reflection using hot mirror films with simplifying assumptions, and his results showed an increase in overall efficiency over a year by 4%. Other efforts of this type used a three-dimension model to take into account the radiation exchange by different segmented surfaces inside the receiver along the pipe's length. This study showed that the hot mirror receiver effectively reduced the IR losses at higher temperatures, reduced the thermal stress on the glass cover and is suggested to be used in a hybrid system [11].
Another system that utilize the IR reflection mechanism inside the annulus of a two concentric cylindrical tubes is the blackbody cavity object. The blackbody is an ideal object, absorbs all incident radiation, regardless of direction and wavelength [12]. The object that most closely resembles a blackbody is a large cavity with a small opening. The radiation that is incident through the opening has very little chance to escape, it is either absorbed or undergoes multiple reflections before being absorbed [12]. Different types of IR reflected coatings in the inner side of the outer cylinder were studied in a cavity design in [13][14][15], where the cavity design was a mirrored cavity receiver with a hot mirror application on the aperture. It showed that the effect of the multiple thermal reflection is able to achieve the highest thermal and optical efficiencies. A good comprehensive review on the topic of the cavity systems is provided by [16].

In this paper, the mathematical model of the IR reflection inside an annulus between two concentric cylindrical tubes for maximum heat retention is discussed in section two. The reflections happened due to the existence of an IR reflected mirror coating in the inside of the outer cylinder. In the third and fourth sections, the numerical model and the simulation implementation of the heat exchange of the system are examined.
In the fifth section, the description of the experiment is presented. Last section is the results of the simulation based on the model that is used to compare experimental results with theoretical expectations. The experimental data is close to theoretical expectations, with results diverging from the simulation by at most 6% at around 700 K, and Chi-squared test p-values around 0.99, with the worst at 0.80. The simulations underestimate the performance, indicating that better performance was measured experimentally.

## 2. Mathematical model

### 2.1. Discretization of the two concentric cylinders

The system of two concentric cylinders is discretized into finite control volumes (CV) to be able to evaluate their thermal interactions. The circumference of the outer cylinder with IR mirror (IRM) and the inner pipe are segmented into $N_l$ CVs along the circumference (azimuthal direction), with $l$ given by $\frac{-(N_l-1)}{2} < l < \frac{(N_l-1)}{2}$ and $N_m$ segments along its axis (axial direction), where $m$ is $\frac{-(N_m-1)}{2} < m < \frac{(N_m-1)}{2}$, see Figure 1.

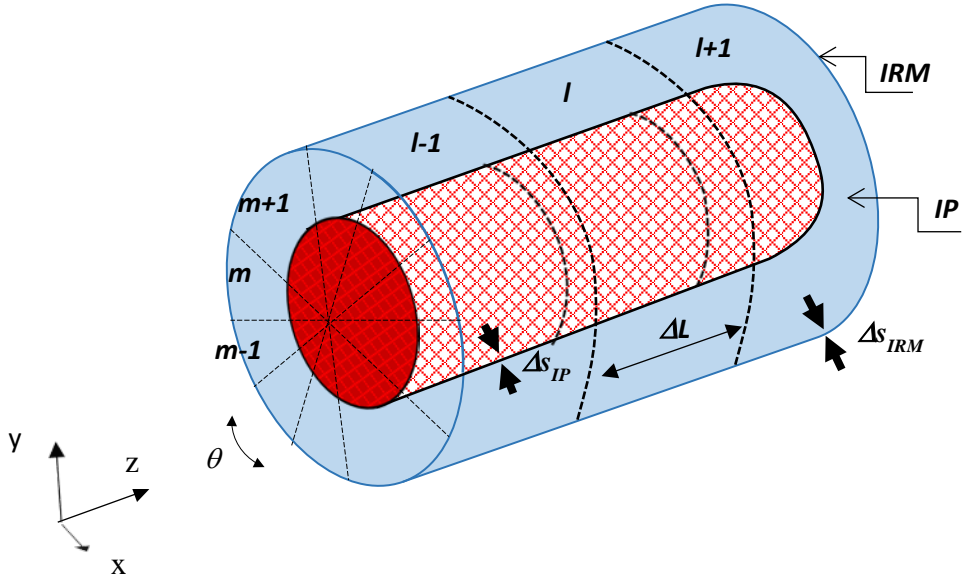

Figure 1: Discretization of the inner pipe (IP) and the outer pipe (IRM) into control volumes.

## 2.2. IR Reflection model

### 2.2.1. The reflected radiation term

An essential aspect of this work is to account for reflected radiation. The reflected radiation is essential when the radiation is reflected onto the Inner Pipe (IP) from the outer pipe that is coated with IR Mirror (IRM) for partial reabsorption. The reflections that we consider are: the first reflection, which is denoted by $(\dot{Q}_{IR,1\,refl})_{ml}$ (represents the reflection from IP to IRM to IP), secondary reflection, which is denoted by $(\dot{Q}_{IR,2\,refl})_{ml}$ (IP to IRM to IP to IRM to IP), and the secondary reflection on the inner cavity surface, which is denoted by $(\dot{Q}_{IR,2\,refl,HM})_{lm}$ (IP to IRM to IP to IRM). The effect of the secondary reflection on IP and IRM depends on the reflection coefficients of both the outer pipe ($\rho_{IRM,IR}$), and the IP ($\rho_{IP,IR}$). We assumed that the optical properties of IRM are approximately constant within the temperature range of interest, and their average values are used. If more detailed results for different temperatures are needed, the parameters can be changed or made temperature dependent on the simulation. We further assumed that the radiation is emitted from the center of the surface of each control volume (IP and IRM). Furthermore, the IP control volumes are diffuse and gray, and the IRM is a specular reflector inside the receiver and opaque from the outside. The IP and IRM optical properties are characterized in terms of visible and IR radiations of the electromagnetic spectrum.

### 2.2.2. The first reflected radiation term $\dot{Q}_{IR,1\,ref}$

The first reflection occurs on the IRM, which is a specularly reflected surface. In Figure 2, $A_{kl}$ control volume on the inner pipe emits IR radiation that is received by another control volume $A_{il}$ on the inner pipe via reflection on the $IRM_{ml}$ control volume. The amount of radiation received by $A_{il}$ control volume depends on the view factor towards the $IRM_{ml}$ control volume. It requires the magnitude of the emitted radiation by $A_{kl}$, reflected radiation by $IRM_{ml}$ and the absorbed radiation by $A_{il}$. Determining the amount of radiation received by $A_{il}$ on the inner pipe from $A_{kl}$, as shown in Figure 2, helps to sum the IR contributions from all the inner pipe control volumes that are in radiative contact with $A_{il}$ via the first reflection. The first reflection can be determined by the help of the simplified scheme in Figure 2. $A_{kl}$ emits radiation diffusely towards the IRM control volumes, a fraction of this radiation strikes $IRM_{ml}$ control volume with an amount of radiation equal $\dot{Q}_{IR,A_{kl}} F_{A_{kl},IRM_{ml}}$, where $\dot{Q}_{IR,A_{kl}}$ is the emitted radiation from $A_{kl}$ and $F_{A_{kl},IRM_{ml}}$ is the view factor from $A_{kl}$ to $IRM_{ml}$. This radiation is going to be received by $A_{il}$ via specular reflection taking place by $IRM_{ml}$ with an amount of radiation equal $\rho_{IRM} \dot{Q}_{IR,A_{kl}} F_{A_{kl},IRM_{ml}}$, where $\rho_{IRM}$ is the reflection coefficient of the IRM.

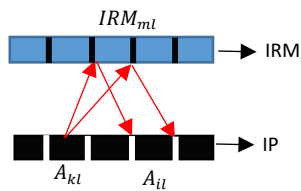

Figure 2: Simple schematic representation to show the mechanism of the first reflection between the inner pipe (IP) and outer pipe (IRM).

As discussed in section 2.1, the system of two concentric cylinders is discretized into control volumes, as shown in Figure 1. The arc length of the inner pipe and IRM control volume have the same central angle "θ". In previous discretization, the control volume "ab" and "cd" were used (Figure 3). Radiation emitted from the center of inner pipe control volume "cd" in a cone is shown in Figure 3. We redefine the outer cover (IRM) control volume such that it lies between normal 1 and normal 2 in order for all radiation from cd to be reflected back onto itself. Geometric optics shows that the arc length between "a" and normal 2 (as well as normal 1 and "b") is exactly half the arc length between normal 1 and normal 2. Therefore, we must increase the number of IRM control volumes to be twice that of the inner pipe control volumes in order to capture all the reflected radiation.

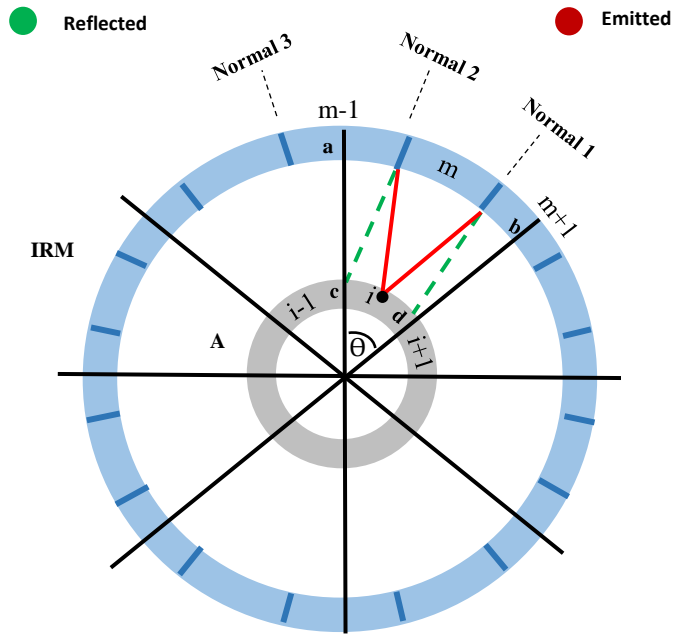

Figure 3: Discretize the IRMM into normal CVs (the index in the axial direction is "l", see Figure 1).

### 2.2.3. Summation of all first reflection contributions to $A_{il}$

The contributions from all inner pipe control volumes that are in radiative contact with $A_{il}$ via reflection will contribute towards the total received IR radiation by $A_{il}$. These contributions were summed. Along the circumference, the farthest IRM control volume that can be viewed by $A_{il}$ constitutes the extreme reflection from the farthest inner pipe control volume which is depicted and calculated in Appendix I.

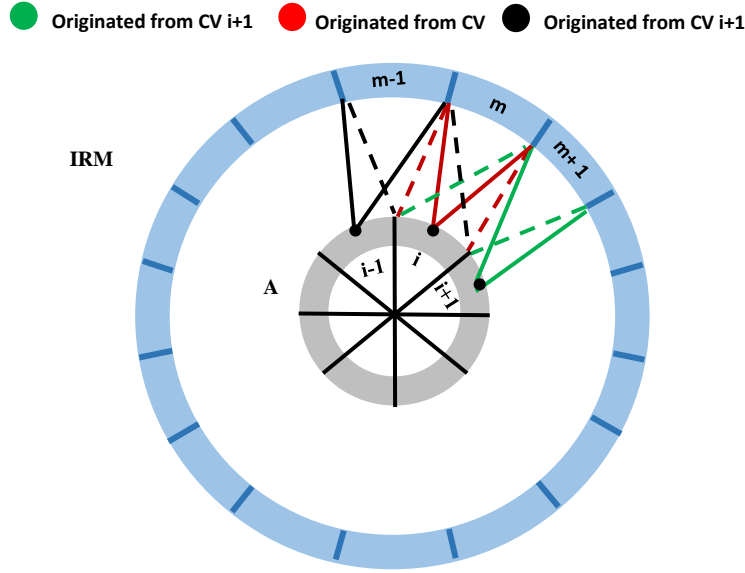

Figure 4: Inner pipe control volume with index $i$ receives reflected radiation from $A_i$, $A_{i+1}$ and $A_{i-1}$ via reflection over $IRM_m$, $IRM_{m+1}$, and $IRM_{m-1}$, respectively.

In Figure 4, the red solid and dash colors represent the emitted radiation from $A_{i,l}$ towards $IRM_{m,l}$ and the reflected radiations from $IRM_{m,l}$ to $A_{i,l}$, respectively. The green solid and dash colors represent the emitted radiation from $A_{i+1,l}$ towards $IRM_{m+1,l}$ and the reflected radiations from $IRM_{m+1,l}$ to $A_{i,l}$, respectively. The black solid and dash colors represent the emitted radiation from $A_{i-1,l}$ towards $IRM_{m-1,l}$ and the reflected radiations from $IRM_{m-1,l}$ to $A_{i,l}$, respectively. The subscripts $i$ and $m$ stand for the index number in the azimuthal direction and $l$ stands for the index in the longitudinal direction. By extending the situation of the first reflection of an emitted radiation via one IRM CV and received by $A_{il}$ (previously mentioned, see Figure 2) to the neighboring elements, see Figure 4, the sum of all contributions on $A_{il}$ by first reflection can be expressed as

$$\left(\dot{Q}_{IR,1ref}\right)_{ml} = \sum_{\substack{k=i+H,\\m=2i+H,\\H=0}}^{H=N_R} \rho_{IRM_{ml}} \dot{Q}_{IR,A_{kl}} F_{A_{k,l}IRM_{m,l}} + \sum_{\substack{k=i-H,\\m=2i-H,\\H=1}}^{H=N_L} \rho_{IRM_{ml}} \dot{Q}_{IR,A_{kl}} F_{A_{k,l}IRM_{m,l}}, \qquad (1)$$

Eq. (1) are the sum of the reflection contributions originating from the right and the left of $A_{il}$, where $N_R$ and $N_L$ are the maximum numbers of inner pipe control volumes that are in radiative contact with $A_{il}$ on their respective sides. $N_R$ and $N_L$ are calculated and discussed in Appendix I.

### 2.2.4. The effect of the first reflection on the IR mirror control volumes

The outer cover with IR mirror (IRM) control volumes partially absorbs the emitted radiation from the inner pipe due to the absorption coefficient of the IRM material. Although the absorption coefficient is very small compared to that of the inner pipe, this contribution was accounted for. In Figure 5 and Figure 6, the two possibilities of light cones originating from different inner pipe control volumes falling onto one IRM control volume are depicted.

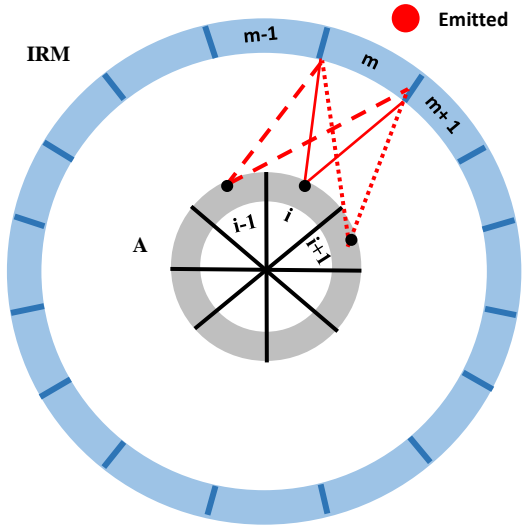 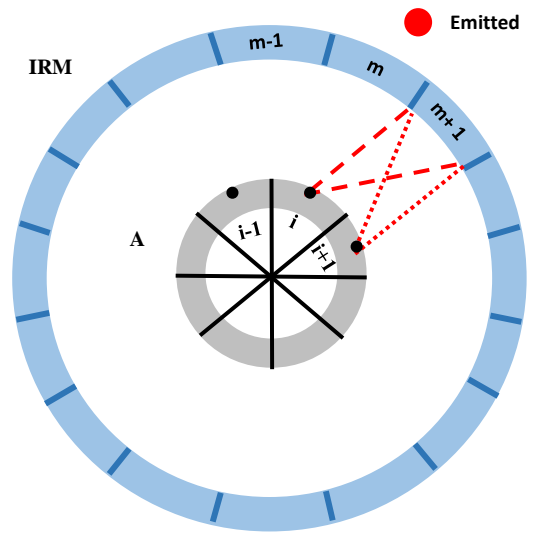

Figure 5: IRM control volume received the emitted inner pipe control volumes radiation (case 1).

Figure 6: IRM control volume received the emitted inner pipe control volumes radiation (case 2).

The absorbed part of the reflected radiation by IRM is represented as follows:

For the case of the IRM control volume in Figure 5

$$\left(\dot{Q}_{IR,1ref,IRM}\right)_{ml} = \left(1 - \rho_{IRM_{ml}}\right) \times \sum_{\substack{k=\left(\frac{m}{2}\right)+H \\ ,H=0}}^{H=N_R} \dot{Q}_{IR,A_{k,l}} F_{A_{k,l}IRM_{m,l}} + \left(1 - \rho_{IRM_{ml}}\right) \times \sum_{\substack{k=\left(\frac{m}{2}\right)-H \\ ,H=1}}^{H=N_L} \dot{Q}_{IR,A_{k,l}} F_{A_{k,l}IRM_{m,l}} \,. \tag{2}$$

For the case of the IRM control volume in Figure 6

$$\left(\dot{Q}_{IR,1ref,IRM}\right)_{ml} = (1 - \rho_{IRM_{ml}}) \times \sum_{\substack{k=\left(\frac{m}{2}\right)+H \\ ,H=1}}^{H=N_R} \dot{Q}_{IR,A_{k,l}} F_{A_{k,l}IRM_{m,l}} + (1 - \rho_{IRM_{ml}}) \times$$
$$\sum_{\substack{k=\left(\frac{m}{2}\right)-H \\ ,H=1}}^{H=N_L} \dot{Q}_{IR,A_{k,l}} F_{A_{k,l}IRM_{m,l}} \ . \tag{3}$$

### 2.2.5. The second reflected radiation term $\dot{Q}_{IR,\,2ref}$

The second reflection is the reflection from the inner pipe (IP) to IRM to IP to IRM to IP. It strongly depends on the values of $\rho_{IRM}$ and $\rho_{IP}$ for the IRM and the inner pipe reflectivity coefficients, respectively. The scenario of the second reflected radiation continues from the first reflection. When an inner pipe control volume receives the reflected radiation, the inner pipe control volume will diffusely reflect the unabsorbed part of this radiation to every direction of its view. These diffusely reflected radiations will again specular reflected by the IRM control volumes towards inner pipe control volumes.

$$\left(\dot{Q}_{IR,2ref}\right)_{il} = \rho_{ab} \sum_{\substack{k=i+H \\ m=2i+H \\ H=0}}^{H=N_R} \rho_{IRM_{m.l}} F_{A_{k,l}IRM_{m,l}} \left(\dot{Q}_{IR,1ref}\right)_{k,l} + \rho_{ab} \sum_{\substack{k=i-H \\ m=2i-H \\ H=1}}^{H=N_L} \rho_{IRM_{m.l}} F_{A_{k,l}IRM_{m,l}} \left(\dot{Q}_{IR,1ref}\right)_{k,l} \tag{4}$$

### 2.2.6. The effect of the second reflection on the IR mirror control volumes

In the same way, as discussed in section 2.2.4, the IRM control volumes partially absorb the diffusely reflected radiations from inner pipe control volumes, but this effect is less than the effect of the first IR reflection on the IRM control volumes. The amount of radiation that has been absorbed by the IRM control volumes due to the second IR reflection is expressed as follows:

For the case of the IRM control volume in Figure 5

$$\left(\dot{Q}_{IR,2ref,IRM}\right)_{ml} = \left(1 - \rho_{IRM_{ml}}\right) \times \rho_{ab} \sum_{\substack{k=\left(\frac{m}{2}\right)+H \\ ,H=0}}^{H=N_R} F_{A_{k,l}IRM_{m,l}} \left(\dot{Q}_{IR,1ref}\right)_{k,l} + \left(1 - \rho_{IRM_{ml}}\right) \times$$
$$\rho_{ab} \sum_{\substack{k=\left(\frac{m}{2}\right)-H \\ ,H=1}}^{H=N_L} F_{A_{k,l}IRM_{m,l}} \left(\dot{Q}_{IR,1ref}\right)_{k,l}.$$
(5)

For the case of the IRM control volume in Figure 6

$$\left(\dot{Q}_{IR,2ref,IRM}\right)_{ml} = \left(1 - \rho_{IRM_{ml}}\right) \times \rho_{ab} \sum_{\substack{k=\left(\frac{m}{2}\right)+H \\ ,H=1}}^{H=N_R} F_{A_{k,l}IRM_{m,l}} \left(\dot{Q}_{IR,1ref}\right)_{k,l} + \left(1 - \rho_{IRM_{ml}}\right) \times$$
$$\rho_{ab} \sum_{\substack{k=\left(\frac{m}{2}\right)-H \\ ,H=1}}^{H=N_L} F_{A_{k,l}IRM_{m,l}} \left(\dot{Q}_{IR,1ref}\right)_{k,l}.$$
(6)

## 3. Numerical model

A model was constructed that describes the different heat transfer interactions inside the system, allowing for including IR reflections inside the annulus between two cylinders. In this section, this model is briefly described. More detail can be found in [11][17]. The physical basis for our model uses energy conservation for the thermal interactions in the system as shown in Figure 7.

The source of the thermal energy used during the experiment was an electrical resistance heater wire with a rate of heat generation which was controlled by a variac. These heating elements are placed inside the inner pipe to bring its surface temperature to desired values. The experiment description is discussed in section 5. The thermal interactions of the system are displayed in a quarter of the cross-sectional view in Figure 7.

Under steady operating conditions, the inner pipe/cylinder and the outer pipe (IRM) reach different stagnation temperatures. Moreover, the heat loss and the heat gain of each element in the system must equal the total rate of heat generation of the heating elements $\dot{E}_{gen}$ (Eq. (7))

$$\dot{q}_{IRM,amb} = \dot{q}_{IRM,cond} = \dot{q}_{IP,IRM} = \dot{q}_{IRM,IP} = \dot{E}_{gen},$$
(7)

where $\dot{q}_{IRM,amb}$ is the rate of the heat transfer from the IRM cover to the surroundings, $\dot{q}_{IRM,cond}$ is the conduction through the IRM layer, $\dot{q}_{IP,IRM}$ is the heat transfer from IP to IRM.

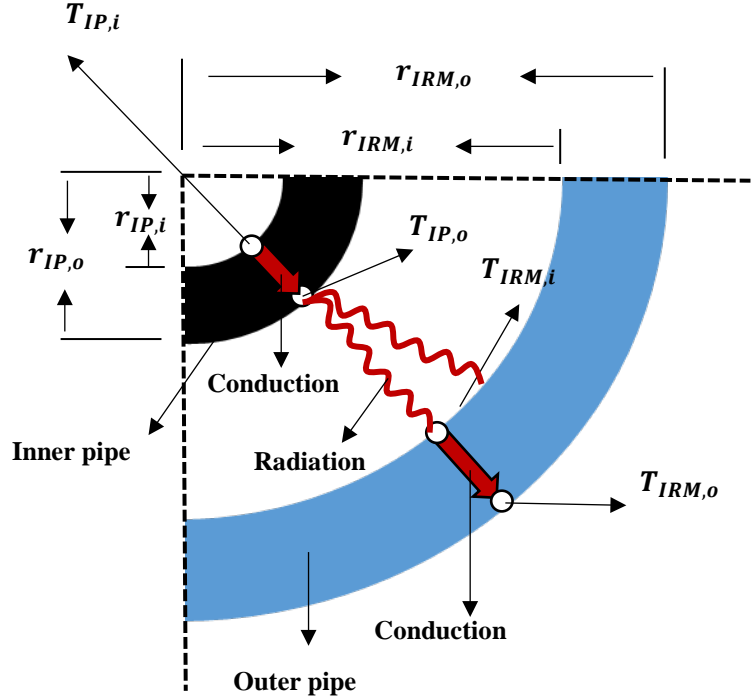

Figure 7: The system cross-section.

The calculations start from the heat loss to the ambient because the ambient temperature is known. We guess the unknown outer IRM surface temperature $T_{IRM,o}$ iteratively, until the steady operating condition at which $\dot{q}_{IRM,amb} = \dot{E}_{gen}$ is fulfilled. The heat rate $\dot{q}_{IRM,amb}$ consists of natural convection and radiation heat transfer from the IRM to the ambient. The air properties during the calculation were selected at $T_{avg} = \frac{T_{IRM,o} + T_{amb}}{2}$.

We can then evaluate $T_{IRM,i}$ at which the rate of heat loss due to the conduction through IRM equals $\dot{E}_{gen}$. In the same way, $T_{IP,o}$ is evaluated through iteration until fulfilling $\dot{q}_{IP,IRM} = \dot{E}_{gen}$, where $\dot{q}_{IP,IRM}$ consists of the rate of the heat transfer between the IP and IRM by convection and radiation. The convection heat transfer inside the evacuated annulus was ignored. The algorithm of solving the above mentioned is discussed in detail in section 4.

## 4. Simulation implementation

Eq. (7) is a set of nonlinear equations which can be linearized via Taylor expansion [18]. The linearized equations become

$$a_m T_m = a_{m+1} T_{m+1} + a_{m-1} T_{m-1} + b_m, \qquad (8)$$

where $T_m$, $a_m$, and $b_m$ are the temperature, the discretization coefficient, and the discretization source term of the control volume of interest, respectively. The remaining terms with index $m+1$ and $m-1$ describe neighboring control volumes. More detail can be found in [11][17].

The algorithm schematic is shown in Figure 8, and is summarized as follows: It starts by guessing the temperature of the outer pipe control volumes (IRM) with a specified initial temperature $T_m^n$. It is an iterative solution until the heat loss of the outer pipe to the ambient is equal to the heat power generation of the heating elements. By knowing the outer pipe temperature, the inner temperature of the outer pipe control volumes (CVs) can be calculated through the conduction heat transfer between outer cover widths. At this point, all control volumes of the same surface are in thermal equilibrium, while the temperatures of different surfaces are different. Next, all IP control volumes obtain a guessed initial temperature $T^*$. The program iterates $T^*$ by solving Eq. (8) for all the control volumes on IP and we obtain a new temperature $T$, which serves as the next guess for next $T^*$ until the convergence criterion, $\frac{T^{n+1}-T^n}{T^n} < 10^{-4}$ ($n$ denotes the $n^{th}$ iteration) is achieved. This process repeats itself for the entire length of the receiver.

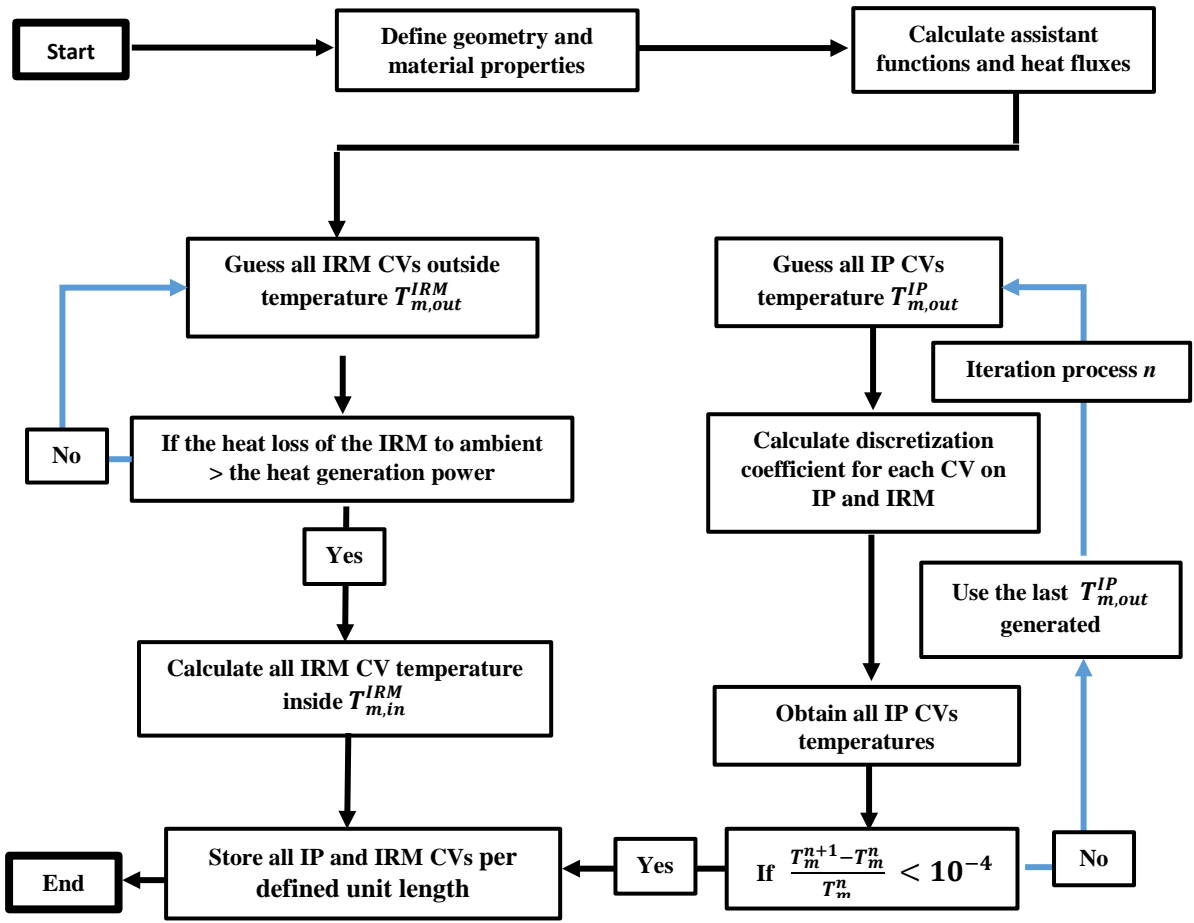

Figure 8: Illustration of the algorithm of the simulation code of the system. Black arrows are for the primary processes execution and the true conditions and the blue arrows, for the false and iterative conditions.

## 5. Experiment descriptions

An experimental setup of two concentric cylindrical tubes was constructed in order to validate the theory and simulation described in the previous section. The unit was constructed as it would be used by heating elements situated inside the inner tube.

The unit was tested indoor. The length of the tested unit was 2.7 m at 25 °C. It consisted of a mild steel inner pipe outer/inner diameter of 0.032/0.028 m and joined pieces of Pyrex glass cover outer/inner diameter of 0.058/0.054 m with a length of 1.35 m each (see Figure 9 and Figure 10). Two such pieces were joined to give a total unit of length 2.70 m. The Pyrex glass pieces were joined with a brass section in the centre of the inner pipe. The central brass piece, glass cover, and the inner pipe were vacuum insulated using flame-resistant high temperature silicon. The air between the inner pipe and the glass envelope was evacuated using an Alcatel vacuum pump (Dual stage rotary vacuum pump input: 208-230 VAC, 60/50 HZ). The high temperature silicon also provided some degree of thermal contact insulation.

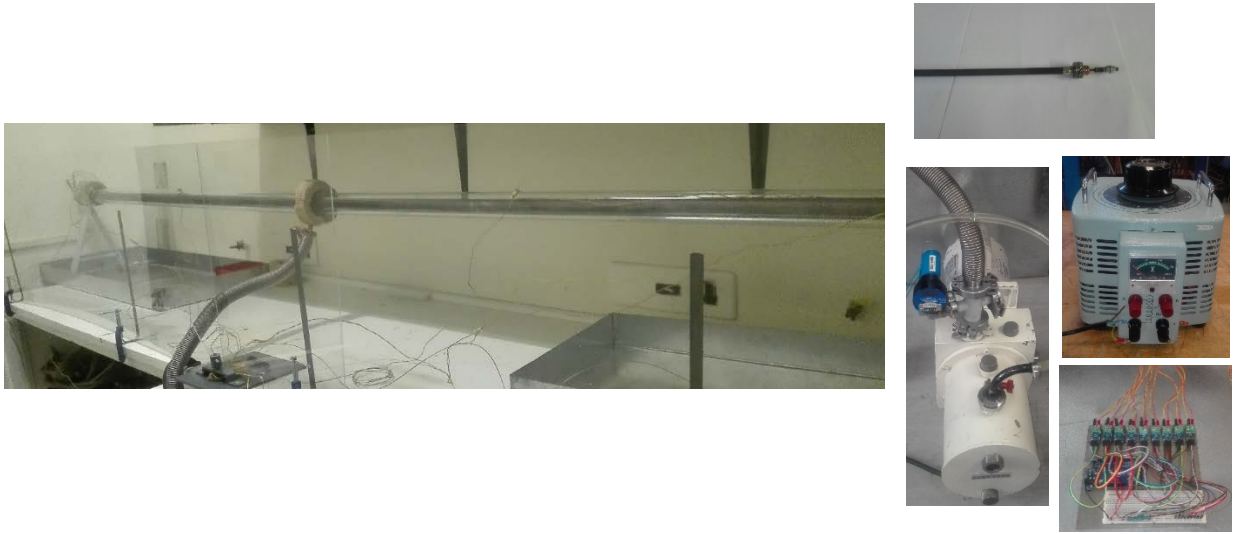

Figure 9: The unit set up in the laboratory, behind a PERSPEX safety shield, on left. On right, top: the heating element. Right center: vacuum pump (left) and variac, and temperature logging unit below.

Two heating elements (1.5 kW tubular element, outer diameter 0.008 m and length 1.172 m, see Figure 9) inside the inner pipe of the unit brought the inner pipe surface temperature to the desired value. The heating element power was adjusted using a variac (Single phase variable transformer, input: 220 VAC 50 Hz output: 0-5 kVAC and 20 A, see Figure 9) and was determined by logging the current and voltage output, with an error of ± 7 W. The heating elements were both 1.2 m in length, with 0.008 m (± $10^{-4}$ m) outer diameter, cold resistance of 16 Ohm (± 0.1 Ohm), and joined electrically inside the inner pipe. In order to prevent the heating element from touching the inner pipe, spacers were introduced to center the heating elements. The inner pipe itself was filled with sand to distribute the heat evenly to the inner pipe surface.

Five thermocouples (K-type thermocouples with a glass fiber twisted insulation, Nickel-Chromium alloy temperature range -200 °C to 1350 °C) were mounted, two on the inner pipe and three on the glass cover, to determine the average temperatures (± 1 °C) and heating behavior of the unit along its' length, see Figure 10. From the temperature information, the heat loss to the environment could be determined.

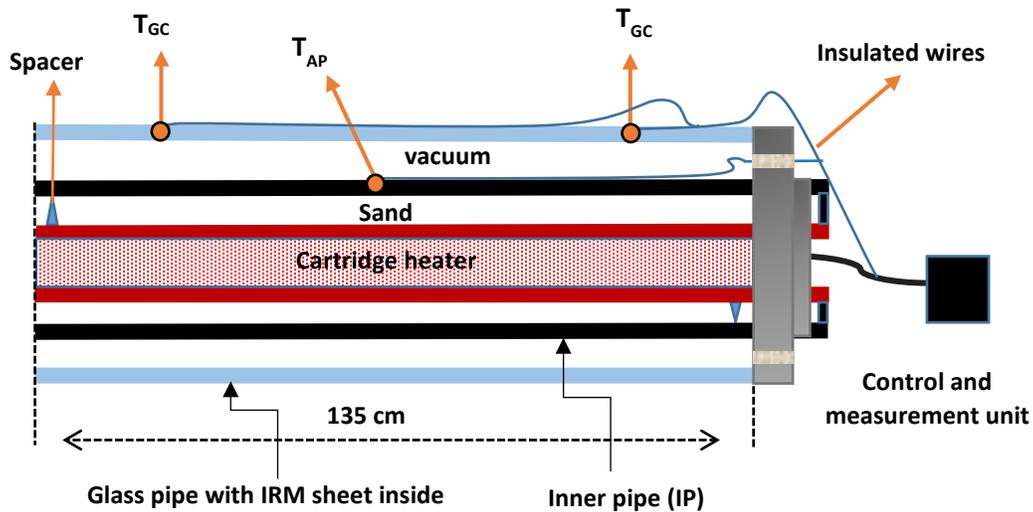

Figure 10: A section of the unit. Glass cover (IRM sheet inside), inner pipe, spacers, heaters, wiring, and thermocouples are shown. "T" represents the thermocouple position.

As indicated in Figure 10, the thermocouples wires were connected to the logging unit (see Figure 9), which allowed regulation and adjustment of the heater power in accordance with temperature requirements.

The experimental procedure was initiated by evacuating the air between the inner pipe and the glass cover, down to a pressure of < 0.1 mbar (measured with a KJL 275i series vacuum gauge), which is sufficient to greatly reduce convective heat transfer in that region [19]. The ambient temperature was noted. Next, a power setting on the variac was chosen, initially around 50 W. A more accurate estimate for the electrical power to the heating elements was determined subsequently using the voltage measurements with a Brymen TBM815 voltmeter (errors on V (AC) = 0.5% & Resistance = 0.1%) from the variac and the temperature-dependent resistance of the heating elements. The temperature measured by the thermocouples was noted using a dedicated microcontroller (Arduino Mega with thermocouple shield (MAX6675)) and displayed on a logging computer. The system was allowed to reach a thermal equilibrium, indicated by a stable reading of inner pipe temperature. The time to reach equilibrium could vary up to about three hours between measurements. Once equilibrium was reached, the temperatures were noted, and the variac setting was increased to the next higher power setting (typically 50 W higher). The experiment was terminated when the power input reached approximately 2 kW, due to concerns of overheating of the vacuum system.

At equilibrium, the electrical power required to maintain the inner pipe temperature equalled the heat loss of the unit through the glass cover. The temperatures along the unit elements were approximately similar to within a few degrees. Heat losses were reported as Power density (Watts per meter) of the receiver unit.

## 6. Results

Two experiments were performed as outlined above. The first used the above system and contained no coating on either the glass cover or the inner pipe, and is designated the name "Bare". This system was investigated because it is the simplest scenario, so any applied coating should perform better in order to be considered. It provided an additional check on our simulation and allowed us to make fine adjustments to the experimental procedure.

The second experiment used an "IRM" coating on the glass cover. A thin, aluminum-based metal sheet with an IR reflectivity of approximately 0.92 was used (obtained from the MIRO SUN Alanod Solar Company datasheet). The sheet is not transparent. The main point of the experiment was to validate the theory and the simulation, with particular emphasis on the IR reflection component, and this must hold for any values.

**The "bare" unit**

The results of the "bare" pipe experiment are displayed in Figure 11. The power density is displayed on the horizontal axis, in units of Watts per meter, and the measured and simulated temperatures are displayed on the vertical axis, in units of degrees Celsius. Both the experimental results (EXP) and the results from the simulation (SIM) are displayed.

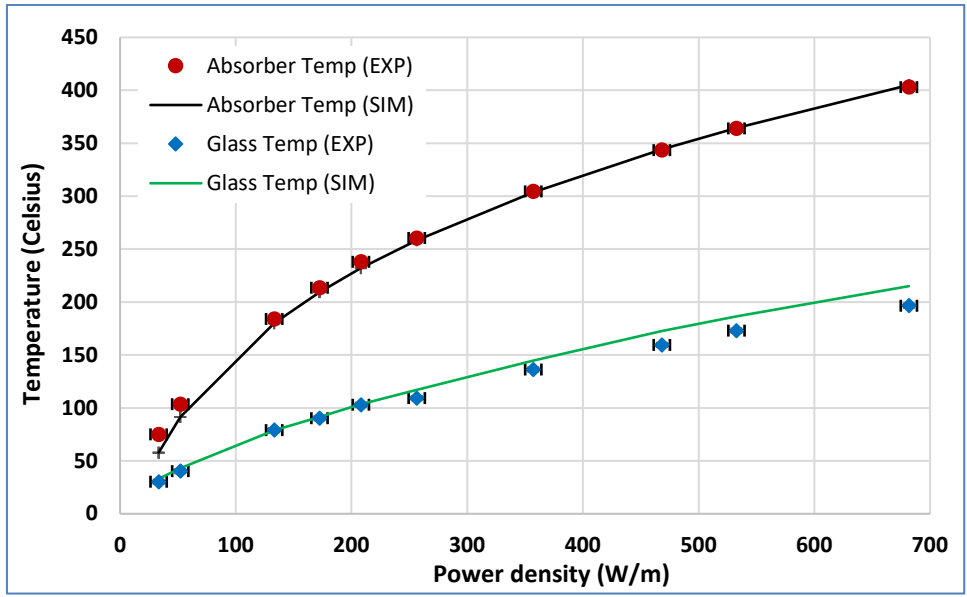

Figure 11: Experimental and simulated results for the temperature profile at different heating powers for a unit without any coating, designated "bare".

A Chi-squared goodness of fit gives p-values of >0.99 for the glass cover and >0.995 for the inner pipe.

**The unit with "IRM" coating**

The results of the second experiment and the simulation are displayed in Figure 12. The axes display the same units as for the "bare" case. It is seen that the experimental (EXP) and simulated (SIM) results for the glass cover with IRM diverge at most by 3%, while those of the inner pipe by at most 6% at around 500 °C, where the simulation underestimates the temperature.

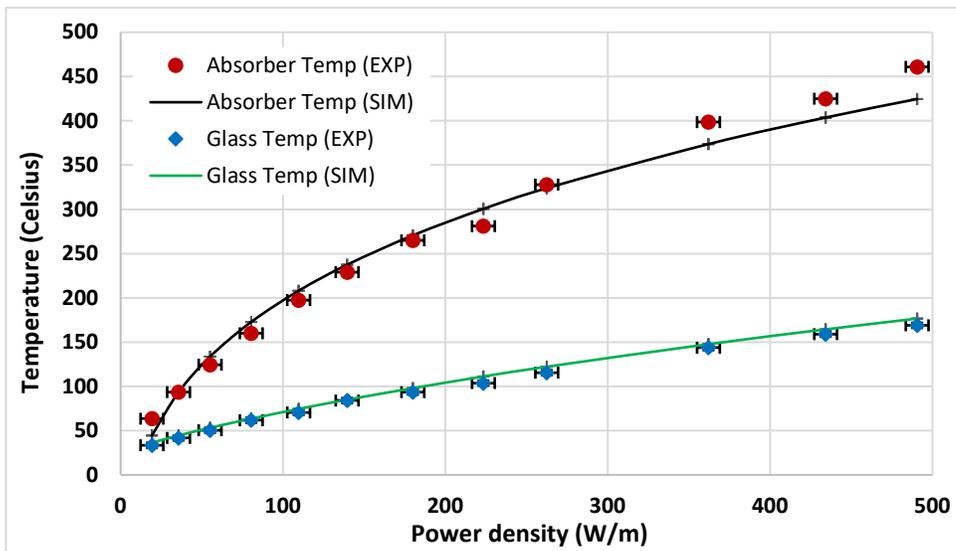

Figure 12: Simulated and experimental results of receiver unit coated with an IRM at various temperatures.

A Chi-squared gives p-values of >0.995 for the glass cover with IRM. For the inner pipe, a p-value of >0.80 was obtained, and the point at 44 °C was considered an outlier so only 10 degrees of freedom were included. The main contribution for the divergence comes from high temperature points, but the simulation underestimates experimental performance. This divergence is likely due to temperature-dependent simulation parameters and can be addressed in a more accurate model.

The effect of the IRM coating can be seen in Figure 13, where the experimental data from the bare inner pipe and the IRM coated outer pipe are displayed.

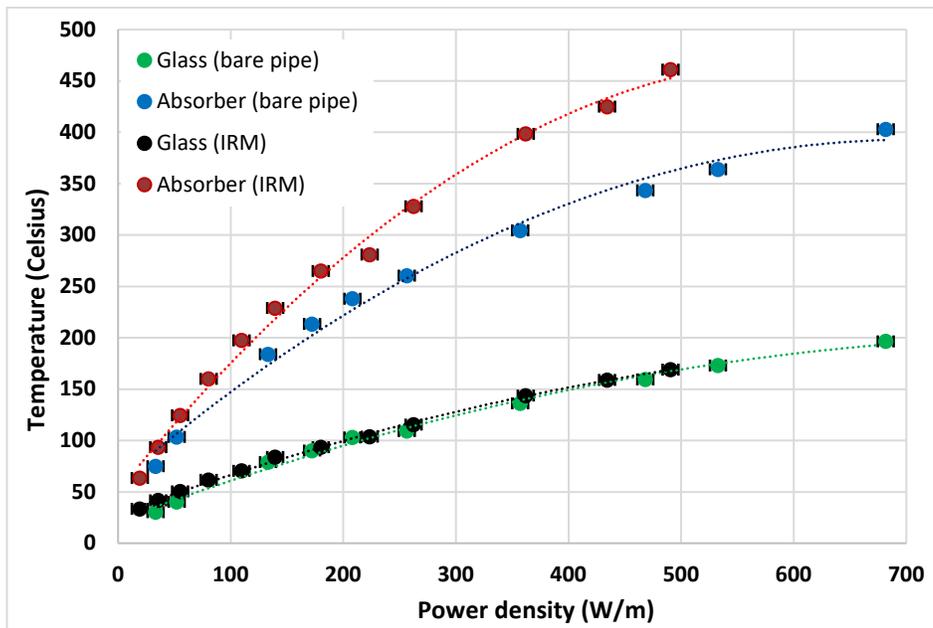

Figure 13: Comparison between experimental results from bare pipe and IRM system. The system with IRM is capable of reaching much higher temperatures at the same power input.

The inner pipe in the IRM case is significantly hotter, indicating better thermal retention, and would, therefore, have a better capability of heating the outer surface of the inner pipe to high temperatures. The outer pipe temperature in both cases is similar, as expected.

## Conclusions

The experimental results from the bare and the application of the IR mirror on the inner side of the outer cover of the system of two concentric cylindrical tubes indicated that the IR reflections inside the system were capable of higher heat retention. Further, the Chi-squared comparison between the experimental and simulated values, which yielded p-values of 0.99 for the glass cover and 0.995 for the inner pipe on the bare pipe, and 0.995 for the glass cover and 0.80 for the inner pipe on the IR mirror application, indicated that the IR reflection model is reasonably accurate to describe an IR reflection inside the annulus of two concentric cylindrical tubes. The simulation underestimated the performance of the IR mirror at higher temperatures. It is likely due to temperature dependent parameters, where the average values in the temperature range of interest were used in the simulation.

**Acknowledgments**

The authors would like to thank the following entities for their kind support: MERG (Materials for Energy Research Group) and MITP (Mandelstam Institute for Theoretical Research), at the University of the Witwatersrand.

**References**

[1]     Porta FL. Technical and economical analysis of future perspectives of solar thermal power plants. *Rep IER* 2005; 1–86.

[2]     Cohen Gilbert E, Kearney DW, Kolb GJ. *Final report on the operation and maintenance improvement program for concentrating solar power plants*. Sandia National Laboratories (SNL), Albuquerque, NM, and Livermore, CA, 1999.

[3]     Fuqiang W, Ziming C. Progress in concentrated solar power technology with parabolic trough collector system: A comprehensive review. *Renew Sustain Energy Rev* 2017; 79: 1314–1328.

[4]     Abdulhamed AJ, Adam NM, Ab-Kadir MZA, et al. Review of solar parabolic-trough collector geometrical and thermal analyses, performance, and applications. *Renew Sustain Energy Rev* 2018; 91: 822–831.

[5]     Granqvist CG. Spectrally Selective Coatings for Energy Efficiency and Solar Applications. *Phys Scr* 1985; 32: 401.

[6]     Canan K. Performance analysis of a novel concentrating photovoltaic combined system. *Energy Convers Manag* 2013; 67: 186–196.

[7]     Miller DC, Khonkar HI, Herrero R, et al. An end of service life assessment of PMMA lenses from veteran concentrator photovoltaic systems. *Sol Energy Mater Sol Cells* 2017; 167: 7–21.

[8]     Lampert CM. Coatings for enhanced photothermal energy collection II. Non-selective and energy control films. *Sol Energy Mater* 1979; 2: 1–17.

[9]     Grena R. Efficiency Gain of a Solar Trough Collector Due to an IR-Reflective Film on the Non-Irradiated Part of the Receiver. *Int J Green Energy* 2011; 8: 715–733.

[10]    Granqvist CG. Radiative heating and cooling with spectrally selective surfaces. *Appl Opt* 1981; 20: 2606–2615.

[11]     Kaluba VS, Ferrer P. A model for hot mirror coating on solar parabolic trough receivers. *J Renew Sustain Energy* 2016; 8: 053703.

[12]     Yunus CA, Afshin JG. *Heat and mass transfer: fundamentals and applications*. 5 Ed. 2015.

[13]     Mohamad K, Ferrer P. Parabolic trough efficiency gain through use of a cavity absorber with a hot mirror. *Appl Energy* 2019; 238: 1250–1257.

[14]     Mohamad K, Ferrer P. Experimental and numerical measurement of the thermal performance for parabolic trough solar concentrators. *Proc 63th Annu Conf South Afr Inst Phys SAIP2018 ArXiv190800515 Physicsapp-Ph*.

[15]     Mohamad K, Ferrer P. Computational comparison of a novel cavity absorber for parabolic trough solar concentrators. *Proc 62th Annu Conf South Afr Inst Phys SAIP2017 ArXiv190713066 Physicsapp-Ph*.

[16]     Mohamad K, Ferrer P. Cavity receiver designs for parabolic trough collector. *ArXiv190911053 Phys*, http://arxiv.org/abs/1909.11053 (2019, accessed 26 September 2019).

[17]     Kaluba VS, Mohamad K, Ferrer P. Experimental and simulated Performance of Heat Mirror Coatings in a Parabolic Trough Receiver. *Submitt Appl Energy J ArXiv190800866 Physicsapp-Ph*.

[18]     Patankar S. *Numerical heat transfer and fluid flow*. CRC press, 1980.

[19]     Forristall R. *Heat transfer analysis and modeling of a parabolic trough solar receiver implemented in engineering equation solver*. National Renewable Energy Lab., Golden, CO.(US), http://large.stanford.edu/publications/coal/references/troughnet/solarfield/docs/34169.pdf (2003, accessed 13 August 2017).

[20]     Hottel HC. *Radiant heat transmission*. Third edition., W.H. McAdams, (Ed.). New York: McGraw-Hill, 1954.

# Appendix I

# The view factor calculations for the system of two concentric cylindrical tubes

The view factor are calculated using the Hottel's crossed string method [20]. To demonstrate this method in the case of a system of two concentric cylindrical tubes, we start with a discretized "control volumes" cross-section of the system, which is shown in Figure I.1

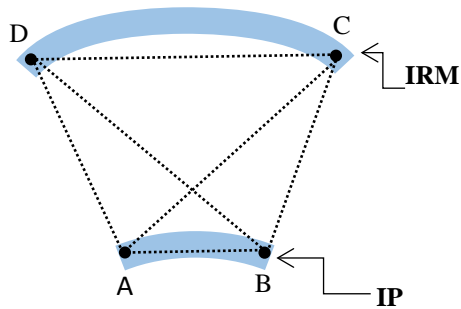

Figure I.1. Determination of the view factor between IP CV and IRM CV using the crossed string method.

According to this method, the view factor from Inner Pipe (IP) to IR Mirror (IRM) equals

$$F_{IP,IRMM} = \frac{\sum crossed\ strings - \sum uncrossed\ strings}{2 \times the\ string\ of\ IP\ surface} = \frac{(AC+BD)-(AD+BC)}{2 \times AB}, \qquad (I.1)$$

## 1. The view factor from IP to IRM

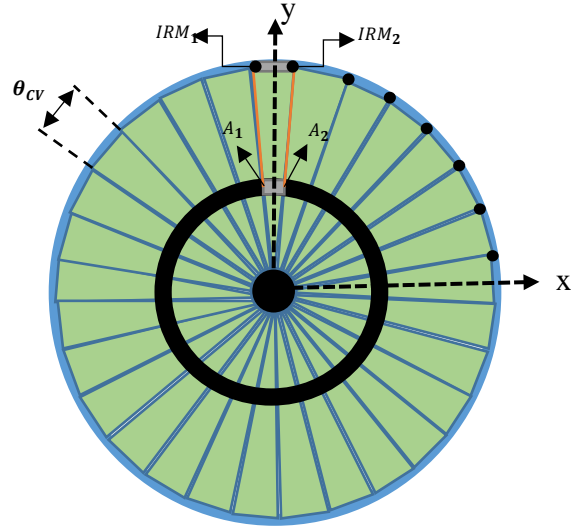

Figure I.2. Description of the view factor between the IRM and inner pipe control volume in a cross-section of the unit.

According to Eq. I.1 and by applying it to the situation in Figure I.2, the view factor from the IRM control volume to IP control volume can be written as

$$F_{A_1A_2,IRM_1IRM_2} = \frac{(|IRM_1-A_2|+|IRM_2-A_1|)-(|IRM_1-A_1|+|IRM_2-A_2|)}{2\times r_{IP,o}\times \theta_{CV}}, \tag{I.2}$$

where $r_{IP,o}$ is the outer radius of the IP and the rest of symbols are shown in Figure I.2. The Cartesian coordinates of the points $IRM_1$, $IRM_2$, $A_1$, and $A_2$ can be easily determined by first obtaining their polar coordinates then write their X and Y components for each point. For example, $IRM_1 = \left(r_{IRM,in}, \left(90 + \frac{\theta_{CV}}{2}\right)\right) = r_{IRM,in}\cos\left(90 + \frac{\theta_{CV}}{2}\right), r_{IRM,in}\sin\left(90 + \frac{\theta_{CV}}{2}\right)$, where $r_{IRM,in}$ is the inner radius of the outer cover and the angle $\left(90 + \frac{\theta_{CV}}{2}\right)$ is measured with respect to the positive X-axis. The view factor from IP control volume ($A_1\,A_2$) to the IRM control volumes can be written as

$$\sum_{ij} F_{A_1A_2,IRM_iIRM_j} = \frac{(|IRM_i-A_2|+|IRM_j-A_1|)-(|IRM_i-A_1|+|IRM_j-A_2|)}{2\times r_{IP,o}\times \theta_{CV}}. \tag{I.3}$$

The summation in Eq. (I.3) continues until the $\overline{IRM_jA_2}$ makes an angle $0°$ with respect X-axis. To obtain the view factor from the IRM control volume to the IP control volumes, we can apply the reciprocity relation

$$A_{IP,CV} F_{A_1A_2, IRM_1IRM_2} = A_{IRM,CV} F_{IRM_1IRM_2, A_1A_2}, \tag{I.4}$$

$$F_{IRM_1IRM_2, A_1A_2} = \frac{A_{IP,CV}}{A_{IRM,CV}} F_{A_1A_2, IRM_1IRM_2}, \tag{I.5}$$

where $A_{IP,CV}$ and $A_{IRM,CV}$ are the surface area of the inner control volume and the IRM cover control volume. Also, the methods mentioned above will provide us with the number of the IRM control volumes that are going to be in thermal contact with IP control volume and vice versa. There is also another method to confirm these number of the IRM control volumes or IP control volumes that are in thermal contact with each other. It will be discussed in the following sections.

**1.1. The number of the IRM segments that are in thermal contact with the inner pipe segment**

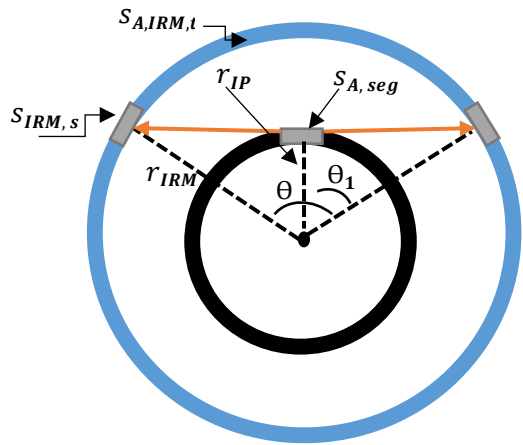

Figure I.3: Graphical representation for the maximum reflection from IP to IRM

In this derivation, we need to evaluate the maximum number of control volume on the IRM cover that is going to be hit by one irradiated IP control volume, as shown in Figure I.3. Starting this derivation by evaluating $\theta_1$

$$\cos \theta_1 = \frac{r_{IP}}{r_{IRM}}, \qquad (I.6)$$

and

$$\theta = 2\theta_1 = 2\cos^{-1}\frac{r_{IP}}{r_{IRM}}. \qquad (I.7)$$

The arc length of the IRM cover control volume is $S_{IRM,seg} = \theta_{seg}\, r_{IRM}$ and the total arc length with a central angle $\theta$ is

$S_{A,IRM,tot} = \theta\, r_{IRM}$. Now, the number of IRM cover segments (control volume's number) that can be affected due to thermal radiation of one IP segment is

$$N^{CV}_{IP-IRMM} = \frac{S_{A,IRM,tot}}{S_{IRM,seg}} = \frac{\theta\, r_{IRM}}{\theta_{seg}\, r_{IRM}} = \frac{2\cos^{-1}(r_{IP}/r_{IRM})}{\theta_{seg}}. \qquad (I.8)$$

**1.2. The number of the inner pipe segments that are in thermal contact with the IRM segment**

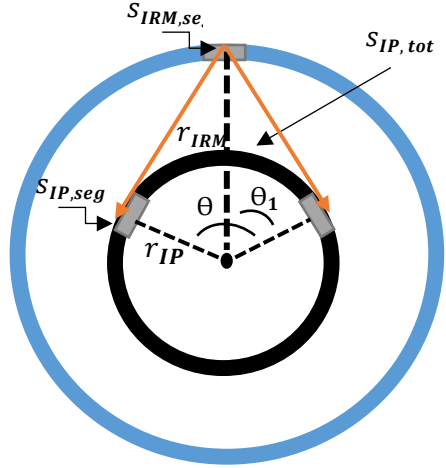

Figure I.4: Graphical representation for the maximum reflection from IRM to IP

This derivation aims to find the maximum number of IP control volumes that are in radiative contact with one IRM control volume. This is due to the reflection from the IRM control volume, as shown in Figure I.4. The derivation starts by evaluating $\theta$, where

$$\theta = 2\theta_1 = 2\cos^{-1}\frac{r_{IP}}{r_{IRM}}. \tag{I.9}$$

By knowing the total arc length of the inner pipe $S_{IP,tot} = \theta\, r_{IP}$ and the arc length of one IP control volume, $S_{IP,seg}$, we can evaluate the number of IP control volumes with a central angle $\theta$ as

$$N_{IRM-IP}^{CV} = \frac{S_{IP,tot}}{S_{IP,seg}} = \frac{\theta\, r_{IP}}{\theta_{seg}\, r_{IP}} = \frac{2\cos^{-1}(r_{IP}/r_{IRM})}{\theta_{seg}}. \tag{I.10}$$

**2. The view factor from IRM to IRM**

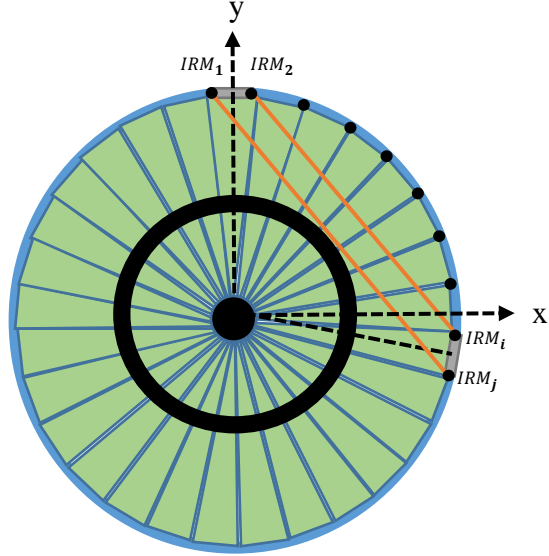

Figure I.5. Description of the view factor from the IRM control volume to IRM control volume in a cross-section of the receiver unit.

Similar to the above section. By using Eq. I.1 for the situation in Figure I.5, the view factor from the IRM control volume to IRM control volume can be written as

$$\sum_{ij} F_{A_1 A_2, IRM_i IRM_j} = \frac{(|IRM_1 - IRM_i| + |IRM_2 - IRM_j|) - (|IRM_1 - IRM_j| + |IRM_2 - IRM_i|)}{2 \times r_{IRM,in} \times \theta_{CV}}, \qquad (I.11)$$

where $r_{IRM,in}$ is the inner radius of the IRM cover, and the other symbols are shown in Fig. I.5. The summation in Eq. (I.11) continues until the $IRM_i$ makes an angle $0°$ with respect to X-axis. This method will help us to get the number of IRM control volumes that are going to be in thermal contact with one IRM control volume. There is also, another way to confirm these number of the IRM control volumes, which is discussed in the following section.

## 2.1. The number of the IRM segments or control volumes (CVs) that are in thermal contact with other IRM segment

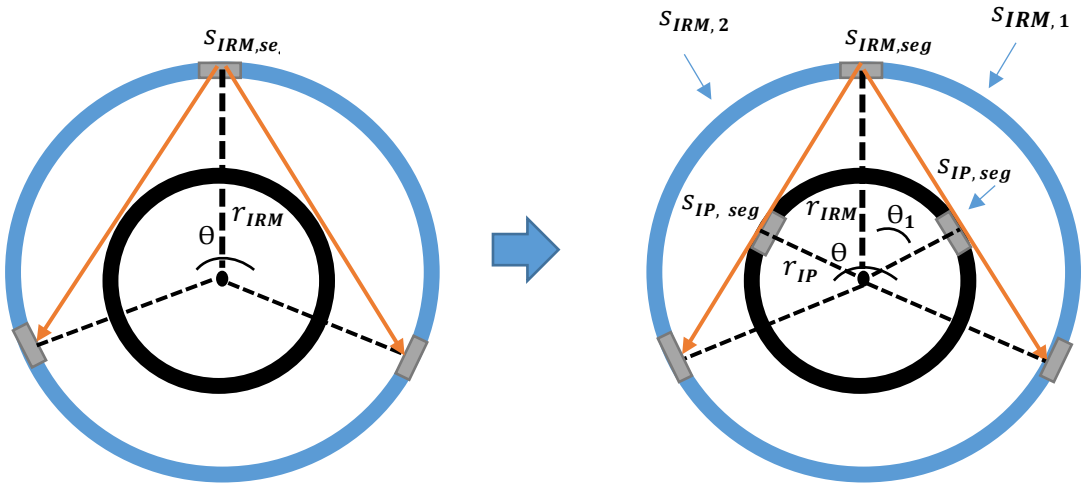

Figure I.6: Graphical representation for the maximum reflection from IRM control volume to IRM control volume

The real representation of the maximum radiation reflected from IRM control volume to IRM control volume is shown on the left of Figure I.6, the extreme rays from IRM control volume to IRM control volume should not touch the IP. If this reflected radiation touches the IP in its way instead of IRM control volume, this reflection will count as an interaction from IP to IRM or vice versa. On the right of Figure I.6, the extreme rays are allowed to touch the IP for the purpose of simplifying the calculations. The calculation starts by evaluating $\theta$, where $\theta = 4\theta_1$ and $\theta_1 = \cos^{-1}\frac{r_{IP}}{r_{IRM}}$.

The total arc length that has the central angle $\theta$ equals $s_{IRM,1} + s_{IRM,2}$. The number of IRM control volumes segments that can be affected due to the thermal radiation from other IRM control volume segment =

$$\frac{s_{IRM,1}+s_{IRM,2}}{s_{IRM,seg}} = \frac{4\,\theta_1\,r_{IRM}}{\theta_{seg}\,r_{IRM}} = \frac{4\cos^{-1}\frac{r_{IP}}{r_{IRM}}}{\theta_{seg}}.$$